# Extreme in-plane upper critical magnetic fields of heavily doped quasi –two-dimensional transition metal dichalcogenides

P. Samuely,[1,2] P. Szabó,[1] J. Kačmarčík,[1] A. Meerschaut,[3] L. Cario,[3] A. G. M. Jansen[4,5], T. Cren,[6] M. Kuzmiak,[1] O. Šofranko[1,2], T. Samuely[2]

[1]*Centre of Low Temperature Physics, Institute of Experimental Physics, Slovak Academy of Sciences, 04001 Košice, Slovakia*
[2]*Centre of Low Temperature Physics, Faculty of Science, P. J. Šafárik University, 04001 Košice, Slovakia*
[3]*Institut des Matériaux Jean Rouxel, Université de Nantes and CNRS-UMR 6502, Nantes 44322, France*
[4]*Universté Grenoble Alpes, CEA, Grenoble INP, IRIG, PHELIQS, F-38000 Grenoble, France*
[5]*Laboratoire National des Champs Magnétiques Intenses (LNCMI-EMFL), CNRS, UGA, F-38042 Grenoble, France*
[6]*Institut des NanoSciences de Paris, Sorbonne Université and CNRS-UMR 7588, Paris 75005, France*

Extreme in-plane upper critical magnetic fields $B_{c2//ab}$ strongly violating the Pauli paramagnetic limit have been observed in the misfit layer $(LaSe)_{1.14}(NbSe_2)$ and $(LaSe)_{1.14}(NbSe_2)_2$ single crystals with $T_c$ = 1.23 K and 5.7 K, respectively. The crystals show a two-dimensional to three-dimensional transition at the temperatures slightly below $T_c$ with an upturn in the temperature dependence of $B_{c2//ab}$, a temperature dependent huge superconducting anisotropy and a cusp-like behavior of the angular dependence of $B_{c2}$. Both misfits are characterized by a strong charge transfer from LaSe to $NbSe_2$. As shown in our previous work $(LaSe)_{1.14}(NbSe_2)_2$ is electronically equivalent to the highly doped $NbSe_2$ monolayers. Then, the strong upper critical field can be attributed to the Ising coupling recently discovered in atomically thin transition metal dichalcogenides with strong spin-orbit coupling and a lack of inversion symmetry. A very similar behavior is found in $(LaSe)_{1.14}(NbSe_2)$, where the charge transfer is nominally twice as big, which could eventually lead to complete filling of the $NbSe_2$ conduction band and opening superconductivity in LaSe. Whatever the particular superconducting mechanism would be, a common denominator in both misfits is that they behave as a stack of almost decoupled superconducting atomic layers proving that Ising superconductivity can also exist in bulk materials.

Subject Areas: Condensed Matter Physics, Materials Science

[*] Corresponding author. Email: samuely@saske.sk

Recently, a new type of superconducting interaction – the Ising pairing – was discovered in the atomically thin superconductors $MoS_2$ [1] and $NbSe_2$ [2]. The lack of crystal inversion symmetry in monolayer combined with strong spin-orbit coupling leads to an effective spin-orbit magnetic field. This fixes the electron spins out of plane (Ising) with opposite signs for the opposite momenta at $K$ and $K'$ of the hexagonal Brillouin zone. The locking of spin and momentum in the superconducting pairing hinders the spin pair-breaking leading to anomalously high in-plane upper critical fields violating the Pauli limit $B_P$. In $NbSe_2$ [2] the Zeeman spin splitting is realized at fields that exceed the superconducting condensation energy almost seven times. It was also shown that upon increasing the number of $NbSe_2$ atomic layers with the onset of interlayer coupling and the restoring of inversion symmetry, the in-plane critical field becomes smaller than $B_P$, calling into question the application of Ising superconductivity in bulk materials.

Here we show that non-conventional superconductivity is at play in a family of bulk compounds made of stacked $NbSe_2$ and LaSe layers. Our transport and *ac* calorimetry measurements down to millikelvin temperatures and in magnetic fields up to 30 T show that in both misfit layer compounds - $(LaSe)_{1.14}(NbSe_2)$ and $(LaSe)_{1.14}(NbSe_2)_2$ - the in-plane critical field $B_{c2//ab}$ is overcoming the Pauli limiting $B_P$ almost 10 and 5 times, respectively. The superconducting anisotropy $\gamma = B_{c2//ab}/B_{c2//c}$ is very high and temperature-dependent. Moreover, the temperature dependence of $B_{c2//ab}$ displays an upturn close to $T_c$, characteristic of a dimensional crossover in vortex matter from three-dimensional (3D) to 2D upon lowering the temperature. The quasi-2D regime is manifested by a cusp in the angular dependence of $B_{c2}$.

Let us first focus on $(LaSe)_{1.14}(NbSe_2)_2$ which is constituted of trilayers where one quasi-quadratic (Q) LaSe chalcogenide plane is sandwiched between two quasi-hexagonal (H) transition metal dichalcogenides (TMD) $NbSe_2$ layers. LaSe is supposed to be a massive electron donor of $NbSe_2$ layer(s) [3]. Such a system is the ideal platform to test 2D physics in a bulk compound. Our experimental work as well as calculations [4] have demonstrated that despite being bulk, the single crystal $(LaSe)_{1.14}(NbSe_2)_2$ behaves as a doped $NbSe_2$ monolayer with a rigid doping of 0.55–0.6 electrons per Nb atom. This doping level can be explained by an intuitive chemical model where each LaSe unit transfers 1 electron to the $NbSe_2$ layer. As there are two $NbSe_2$ units per 1.14 LaSe unit, one expects a charge transfer of 0.57 electron per $NbSe_2$ unit. This is precisely what our previous scanning tunneling microscope (STM) and angle resolved photoemission spectroscopy (ARPES) measurements and density functional theory (DFT) calculations show. Our work thus confirms that Ising superconducting coupling may be a possible scenario for the bulk compound $(LaSe)_{1.14}(NbSe_2)_2$. Notice that until now only monolayer- or few-layer systems were found to exhibit Ising superconductivity. Our wok suggests that bulk compounds could exhibit it, too.

If we use the same chemical model for $(LaSe)_{1.14}(NbSe_2)$ where LaSe layers are simply alternating with $NbSe_2$ planes, we find a charge transfer of 1.14 electron per $NbSe_2$ unit. However, this is impossible since the undoped $NbSe_2$ unit can accept at most one electron. This indicates that the $NbSe_2$ band could be completely filled, while the metallicity/superconductivity would reside in LaSe layers. On the other hand, vacancies [5] can reduce the charge transfer, keeping the $NbSe_2$ band conducting with much fewer holes than in the case of $(LaSe)_{1.14}(NbSe_2)_2$. A comprehensive study of the superconducting misfit $(LaSe)_{1.14}(NbSe_2)_{m=1,2}$ compounds would therefore be of great interest.

The misfit TMDs [6] have a formula $(MX)_{1+x}(TX_2)_m$ (M=Sn, Sb, Pb, Bi, or Rare-Earth, X = Chalcogen, T = Transition metal, $x$ misfit parameter, and m = 1, 2, 3) indicating alternating stacking of MX planes (having NaCl structure, quasi-quadratic, $Q$) and TMD layers ($CdI_2$ or $NbS_2$ structure, quasi-hexagonal, $H$). Due to different symmetries of the respective layers misfit results along the crystallographic *a* axis. We prepared the misfit layered systems made of $LaSe/NbSe_2$ layers, namely $(LaSe)_{1.14}(NbSe_2)_2$, here denoted as $1Q2H$, which has the highest transition temperature $T_c$ = 5.3 K [7] among TMD misfits and $(LaSe)_{1.14}(NbSe_2)$, denoted as $1Q1H$ with $T_c$ about 1.3 K [8]. Single crystals were prepared by direct reaction of the La, Nb and Se elements in stoichiometric ratios as explained elsewhere [7]. The synthesis yielded large black crystals grown on the surface of a black powder. Energy-dispersive X-ray spectroscopy and X-ray diffraction techniques confirmed the expected compositions and cell parameters of the $1Q1H$ and $1Q2H$ misfit compounds.

A standard lock-in technique was used to measure the magnetic field dependence of the sample resistance in a four-probe configuration at different fixed temperatures down to 100 mK and in magnetic fields up to 30 T in the Laboratoire National de Champs Magnétiques Intenses in Grenoble. The magnetoresistive transitions were also measured at different angles between the *ab* plane of the sample and the applied magnetic field, keeping the current always orthogonal to the field. The angular resolution was better than 0.2 degree with the θ = 0° orientation in the *ab* plane defined by the highest value of $B_{c2}$. The temperature dependence of the resistance at fixed magnetic fields up to 8 T was measured in a He-3 refrigerator in the Centre of Low Temperature Physics Košice. An *ac* calorimeter installed in the same He-3 cryostat was used to measure the temperature and field sweeps of the specific heat of the same sample as for the resistive measurements. *ac* calorimetry [9,10] employs periodically modulated power on the sample and measures the resulting sinusoidal temperature response. In our case, heat is supplied to the sample at a frequency ω ~ several Hz by a light-emitting diode via optical fiber. The chromel-constantan thermocouple calibrated in the magnetic field is used to record the temperature oscillations.

Although *ac* calorimetry is not capable of measuring the absolute values of the heat capacity, it is very sensitive to relative changes in minute samples.

Figure 1a) shows the structure of the $1Q1H$ and $1Q2H$ crystals where quasi hexagonal $NbSe_2$ and quasi quadratic LaSe layers are stacked. The lattice vectors $\vec{a}$, $\vec{b}$ and $\vec{c}$ are indicated.

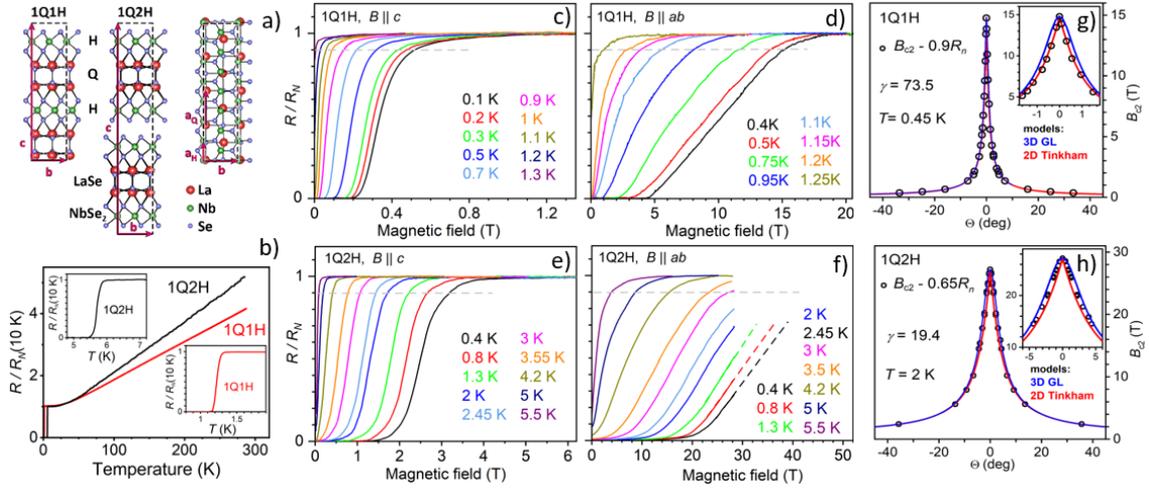

FIGURE 1. Crystal structure with lattice vectors represented by red arrows a) Left – $(LaSe)_{1.14}(NbSe_2)$ / $1Q1H$. Center – $(LaSe)_{1.14}(NbSe_2)_2$ / $1Q2H$. In both cases unit cell is indicated by dashed rectangle. Note that trilayer stacks of $1Q2H$ are decoupled by van der Waals gap. Right – top view, the dashed rectangle represents a nearly commensurate approximate of the crystal structure with $4a_Q \approx 7a_H$. b) Temperature dependence of resistivity normalized to residual value for $1Q2H$ and $1Q1H$, respectively. Insets show superconducting transition in detail. c) and d) Resistive transitions as a function of magnetic field oriented perpendicular (Ref. 8) and parallel to the *ab* plane of the $1Q1H$ misfit compound at fixed temperatures, respectively. e) and f) resistive transitions for the misfit $1Q2H$ for *B* perpendicular and parallel with *ab* planes of the $1Q2H$ system, respectively. Dashed lines in f) extrapolate resistivity measurements up to fields where linear behavior is expected. Short horizontal dashed lines in c) – f) indicate $R/R_N$=0.9, resp. 0.7. g) shows angular dependence $B_{c2}(\theta)$ of the $1Q1H$ sample taken at 0.45 K. h) shows angular dependence $B_{c2}(\theta)$ of the $1Q2H$ sample taken at 2 K. $B_{c2}(\theta)$ dependences are compared with the anisotropic 3D GL model and 2D model of Tinkham. Insets display a zoom.

In the case of $1Q1H$ one has an alternation of LaSe and $NbSe_2$ layers bound by iono-covalent bonding (left). Notice that this is not a van der Waals material. The center of Fig. 1a) shows the structure of the $1Q2H$ compound, a central slab of LaSe/$Q$ is sandwiched between two $NbSe_2$/$H$ layers. The $1Q2H$ trilayer is bound by iono-covalent bonding while the crystal is made by van der Waals stacking of the trilayers. Misfit occurs due to mismatch of the $\vec{a}_Q$ and $\vec{a}_H$ vectors of the LaSe and $NbSe_2$ sublattices, respectively as shown on the right.

Figure 1 b) displays the temperature dependence of the resistance *R* of both $1Q1H$ and $1Q2H$ samples indicating a good metallic conductivity with the residual resistivity ratio RRR = 4.3 and 5.5, respectively. Both samples reveal a sharp transition to the superconducting state with $T_c=$ 1.23 K and the width of the transition between 0.1 and $0.9R_n$ being $\Delta T_c=$ 0.1 K for 1Q1H and $T_c=$ 5.7 K and $\Delta T_c=$ 0.2 K for the 1*Q*2*H* compound.

The resistive transitions from the superconducting to normal state in applied magnetic field oriented parallel to the *c* axis and parallel to the *ab* plane are presented in Figs. 1 c – f) for both compounds. The measured resistances were normalized to the high field values. The transitions are shifted to higher fields and broadened with decreasing temperature. The 1*Q*1*H* sample shows the superconducting transitions at the fields parallel to the *c* axis below 0.5 T down to 100 mK. On the other hand, for the fields parallel to the *ab* plane the onset of the normal state starts at 5 T but the full normal state is achieved only above 18 T at 400 mK. In the case of 1*Q*2*H* the transitions at fields parallel to *c* direction are below 3 T while for the in-plane fields the transition starts at 20 T and goes far beyond the 28 T limit of the Bitter magnet in Grenoble pointing at some 50 T at 400 mK. We determined the upper critical magnetic field at 90 percent of the transition to the normal state $(0.9R_n)$ to better estimate the huge extent of the superconducting state in these low $T_c$ samples. The determination of $B_{c2}$ at the midpoint of the transition yields qualitatively the same conclusions (See Supplemental Material [11]).

To evaluate the temperature dependence of $B_{c2}$ in more detail we made the measurements of the temperature dependence of the resistance at fixed magnetic field in a superconducting coil. The measurements were performed on the 1*Q*1*H* sample from the same batch as before in the Bitter coil and a very good agreement between these two measurements was found. Moreover, on the same piece of the sample we performed specific-heat (*C*) measurements using a very sensitive *ac* calorimetry [9]. By subtracting the normal-state measurement obtained at 1-T field oriented parallel to the c axis from the one in the superconducting state we eliminate contributions from the phonons and from the addenda and obtain a difference of the electronic specific heat $\Delta C/T$. In Fig. 2 a) the superconducting transitions are displayed for the indicated fields parallel to the *c* axis while in Fig. 2 b) they are displayed for fields parallel with the *ab* plane. One can see that at zero field the superconducting anomaly in specific heat characteristic of the second-order phase transition arises close to the temperature where the resistance sharply drops to zero. For the in-field measurements in both field orientations the transitions get broadened, but a very good agreement is found between both, the superconducting onsets and midpoints of the respective resistive transition and the specific-heat anomaly. This strongly suggests that both physical quantities are well characterizing the bulk superconductivity in the system. Remarkably, even if the specific-heat anomaly is well pronounced, the amplitude of the anomaly is quite rapidly suppressed upon increasing magnetic field. This is reminiscent of the situation in high-$T_c$ cuprate superconductors [12,13], where the broadening of the transition in a magnetic field was attributed to the weakening of the coupling between the $CuO_2$ planes that limits the superconducting order to two dimensions, thereby enhancing fluctuations.

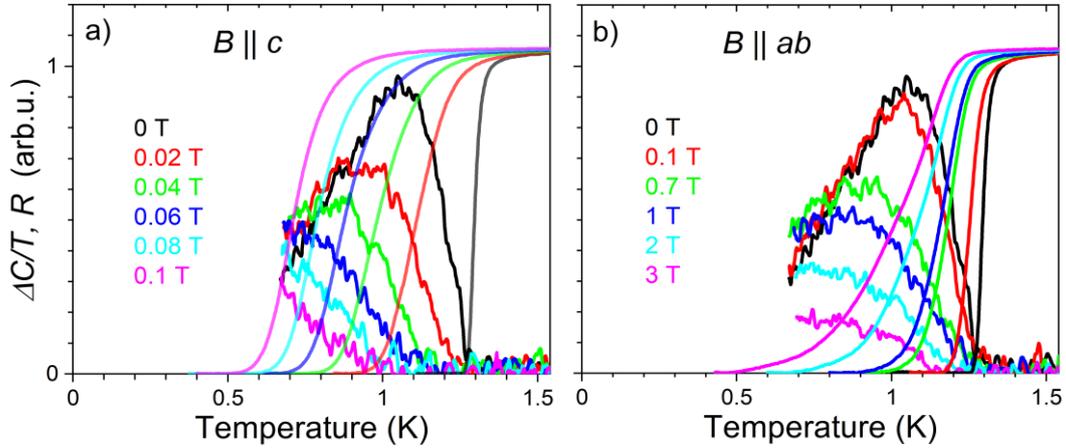

FIGURE 2. Superconducting transitions of $(LaSe)_{1.14}(NbSe_2)$ obtained by resistive and heat capacity measurements. a) the superconducting transitions are displayed for the indicated fields parallel to the *c* axis while in b) for the fields parallel with the *ab* plane.

Figure 3 presents the main result of the paper showing the temperature dependence of upper critical fields for both principal field orientations. Figure 3a) shows the in-plane $B_{c2}$ values of 1*Q*1*H* obtained at 0.9 $R_n$ of $R(B)$ – solid red points and $R(T)$ characteristics – open red points. The black asterisks are obtained at the onset of the specific heat anomaly in Fig. 2b). Below $T_c$ the resulting $B_{c2//ab}(T)$ dependence shows a pronounced upturn and changes a curvature from positive to negative. At lower temperatures $B_{c2//ab}(T)$ can be well approximated by a square-root temperature dependence (dashed line). The experimental value $B_{c2//ab}(0.4\ K) = 15$ T and extrapolated $B_{c2//ab}(0\ K)$ is close to 19 T. The upper limit of the in-plane upper critical fields is obtained from the interlayer transport, namely from the conductance $\sigma_c(B)=1/R_c(B)$ (see Supplemental Material [11]) with $B_{c2//ab}(0\ K)$ above 20 T (gray points and dashed line - approximated square-root temperature dependence). Thus, for fields parallel to the *ab* plane of 1Q1H all the measurements, namely the specific heat, intralayer (in-plane) resistivity and interlayer resistivity provide very consistent results pointing to extremely high upper critical field exceeding the Pauli limit $B_P=2.2$ T (green dashed line in Fig. 3 a)) by a factor of about 9-10, pointing to a large Zeeman-type spin-orbit interaction. Also, the square-root temperature dependence $B_{c2//ab}(T)$ is compatible with the spin pair breaking [14]. Contrarily, the orbital pair breaking is naturally present due to a large coherence length in the configuration when the magnetic field is perpendicular to the ab plane, i.e., in the *c* direction. The blue points and the pink asterisks show the temperature dependence of $B_{c2//c}(T)$ obtained from $R(B)$ in Fig. 1c) and from specific heat in Fig. 2a), respectively. $B_{c2//c}$ values are multiplied by factor 10 for clarity. The temperature dependence shows a positive curvature below $T_c$ changing to a negative one below $T_c/2$ with extrapolated $B_{c2//c}(0\ K)$ close to 0.5 T. Then, the superconducting anisotropy $\gamma = B_{c2//ab}/B_{c2//c}$ at low temperatures reaches about 40.

In Fig. 3 b) the upper critical magnetic fields for 1$Q$2$H$ are presented, constructed in a similar way as for 1$Q$1$H$. Here, $B_{c2//ab}$ established at 0.9 $R_n$ (red points) from the resistive transitions is accessible only down to 3 K where $B_{c2//ab} = 27.3$ T. Then, we also present the transitions taken at 0.7 $R_n$ by the light red points. One can see that below $T_c = 5.7$ K $B_{c2//ab}(T)$ has a positive curvature changing to a negative one below 4 K. The low-temperature $B_{c2//ab}(T)$ could be approximated by the square-root temperature dependence providing $B_{c2//ab}$(0 K) equal to 40 and 53 T at 0.7 and 0.9 $R_n$, respectively. This is 4 to 5 times higher than the Pauli limit $B_P$=10.5 T indicated by the green dashed line. The $B_{c2//c}(T)$ shown by the blue points (values multiplied by 5 for clarity) reveals a positive curvature all the way down to the lowest temperature of 400 mK. The superconducting anisotropy $\gamma = B_{c2//ab}/B_{c2//c}$ at low temperatures is approximately 15.

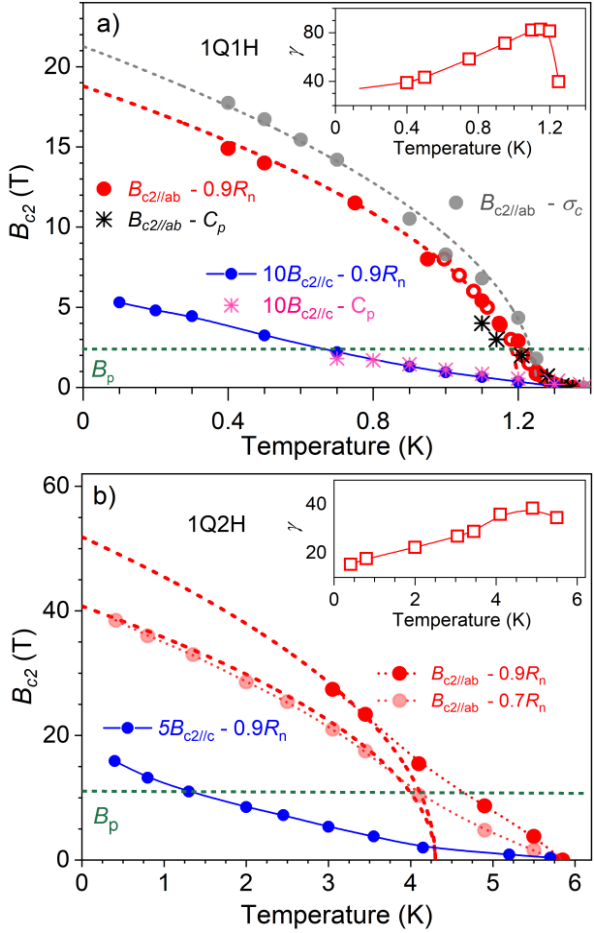

FIGURE 3. Upper critical magnetic fields of a) $(LaSe)_{1.14}(NbSe_2)$ / 1$Q$1$H$ and b) $(LaSe)_{1.14}(NbSe_2)_2$ / 1$Q$2$H$ misfit compounds, red and blue points show $B_{c2}$ for fields applied parallel and perpendicular to ab planes, respectively as determined at the 90% of the resistive transition. In a) solid red points are from field sweeps and open red points from temperature sweeps of the resistance. Gray points are determined from the interlayer transport, namely from the conductance $\sigma_c(B)=1/R_c(B)$ (see Supplemental Material [11]). Stars denote values obtained from the onset of the heat capacity anomaly. Light red points in b) are taken at 0.7 $R_n$. For better readability $B_{c2//c}$ values are multiplied by 10 and 5 for 1$Q$1$H$ and 1$Q$2$H$, respectively. Green dashed lines show Pauli paramagnetic limiting field. In the case of 1$Q$2$H$ $B_{c2//ab}$ for fields larger than 30 T are determined at 70% of the resistive transitions. Red and grey dashed lines are approximations by a square root temperature dependence. Solid and dotted lines are guides for the eye. The temperature dependence of the superconducting anisotropy $\gamma = B_{c2//ab}/B_{c2//c}$ for both compounds are shown in the insets.

2D superconductivity can also be addressed via studying the angular dependence of the upper critical field $B_{c2}(\theta)$, where $\theta$ is the angle between the applied field and the *ab* plane of the sample. Within the anisotropic 3D Ginzburg-Landau (GL) model $B_{c2}(\theta)$ can be described by a simple ellipsoidal formula with a rounded maximum around $\theta = 0$ (magnetic field parallel to the layers) $\left(\frac{B_{c2}(\Theta)\sin\Theta}{B_{c2||c}}\right)^2 + \left(\frac{B_{c2}(\Theta)\cos\Theta}{B_{c2||ab}}\right)^2 = 1$. In 2D regime the superconducting layers are decoupled and can be treated as isolated thin films. For this case Tinkham [15] proposed the equation $\left|\frac{B_{c2}(\Theta)\sin\Theta}{B_{c2||c}}\right| + \left(\frac{B_{c2}(\Theta)\cos\Theta}{B_{c2||ab}}\right)^2 = 1$, with a finite slopes at $\theta = 0$ making a cusp. This effect was also observed in superconducting superlattices below the 2D-3D transition [16] and also in the case of the extremely anisotropic $Bi_{2.2}Sr_{1.8}CaCu_2O_{8+\delta}$ [17].

In Fig. 1 g) we present $B_{c2}(\theta)$ of the 1*Q*1*H* sample taken at 0.45 K. This particular 1*Q*1*H* sample has lower perpendicular critical fields $B_{c2//c}$ than that in Fig. 1 c), probably due to smaller disorder but it has a very similar $B_{c2//ab}$. $B_{c2}(\theta)$ data are compared with both models: the anisotropic 3D GL and 2D model of Tinkham. For such a high anisotropy ($\gamma$=73.5) the difference is minor but the inset showing small angles is more compatible with the 2D model indicating the two-dimensional character of the vortex matter. Figure 1 h) shows the angular dependence $B_{c2}(\theta)$ for the 1*Q*2*H* sample taken at 2 K. Even the inset displaying small angles is not capable of distinguishing the two models since the experimental data lie exactly in between. It is noteworthy that these experiments are rather challenging and $B_{c2}(\theta)$ with cusplike behavior could be observed only on very tiny samples with a clear plane parallel geometry, while in samples with a slightly wavy surfaces the angular dependence always revealed the round maximum, probably due to angle averaging.

Lawrence and Doniach [18] described the behavior of layered superconductors in a magnetic field by a model based on stacked two-dimensional superconducting layers coupled together by Josephson tunneling between adjacent planes. In contrast to the isotropic case where the magnetic flux penetration occurs in the form of vortices of circular symmetry, in anisotropic superconductors the vortex cores will be flattened in the interlayer *c* direction for fields parallel to the layered structure with $\xi_c < \xi_{ab}$ for the vortex core radii $\xi_c$ and $\xi_{ab}$, respectively, perpendicular and parallel to the planes. If $\xi_c$ is bigger than the distance between the adjacent superconducting layers, the system is anisotropic but still three-dimensional with the upper critical magnetic fields for fields perpendicular and parallel to the layers determined by a product of the corresponding coherence lengths: $B_{c2//c} = \Phi_0/(2\pi\xi_{ab}^2)$ and $B_{c2//ab} = \Phi_0/(2\pi\xi_{ab}\xi_c)$, respectively, where $\Phi_0$ is the flux quantum. In some cases, including the high-$T_c$ cuprates, the critical fields can be very high and in extremely anisotropic superconductors $\xi_c$ could shrink with decreasing temperature below the value of the interlayer distance $D$. Then, the vortices become confined between the superconducting layers for parallel fields leading to a dimensional crossover. According to the simplest Lawrence-Doniach model the parallel upper critical $B_{c2//ab}$ is predicted to diverge at the temperature $T^*$, where $\xi_c(T^*) \approx D$. The real finite value of $B_{c2//ab}$ is caused by the finite superconducting layer thickness, Pauli paramagnetism and spin-orbit scattering. Klemm, Luther and Beasley (KLB) [19] have extended the Lawrence-Doniach model to include these effects. They show that the divergence is removed but the dimensional crossover to a two-dimensional superconductivity is still characterized by a strong upward curvature of $B_{c2//ab}(T)$. Experiments on the intercalated layered compounds based on 2$H$-TaS$_2$ [20] revealed a strong upward curvature of the temperature dependence of $B_{c2//ab}$ accompanied by temperature-dependent critical field anisotropy $\gamma = B_{c2//ab}/B_{c2//c}$ reaching maximum values of about 60 at the dimensional crossover and strong in-plane upper critical fields achieving up to three $B_P$ on account of the strong spin-orbit scattering rate due to collisions at the interfaces between TaS$_2$ layers with the heavy Ta atom ($Z$=73) and the light organic layers of intercalants.

In both (LaSe)$_{1.14}$(NbSe$_2$) and (LaSe)$_{1.14}$(NbSe$_2$)$_2$, a dimensional crossover is evidenced by an upturn of $B_{c2//ab}(T)$ and also the temperature dependent anisotropy factor $\gamma$ (see insets in Fig. 3). From $B_{c2//c}$ data we can estimate the size of the coherence length $\xi_{ab}$. For 1$Q$1$H$ with $\xi_{ab}(1.2 \text{ K}) \approx$ 97 nm taking into account the anisotropy factor of about 82 one obtains $\xi_c(1.2 \text{ K}) = 1.2$ nm which is close to the thickness of a stack of one $H$ and one $Q$ layers, 1.2 nm [21]. Similarly, for 1$Q$2$H$ at 5 K we obtain $\xi_{ab} \approx 40$ nm and with $\gamma \approx 40$ we get $\xi_c \approx 1$ nm. Again this is close to the distance of about 1.2 nm separating two NbSe$_2$ layers across a LaSe layer in this compound [7]. Then, the conditions for dimensional transition are fulfilled.

In our previous work [8] we found that the (LaSe)$_{1.14}$(NbSe$_2$) misfit layer compound behaves as a stack of intrinsic Josephson junctions making it a 1-Kelvin analogue to the high-$T_c$ Bi$_2$Sr$_2$CaCu$_2$O$_{8+\delta}$ superconductor [22]. For this to be case, the interlayer superconducting coherence length $\xi_c$ should be shorter than the distance between superconducting atomic layers. In this limit, one can see (LaSe)$_{1.14}$(NbSe$_2$) misfit as slabs of superconducting layers separated by insulating ones behaving as Josephson tunnel junctions. The short $\xi_c$ imposes such a high in-plane upper critical magnetic field $B_{c2//ab} \propto 1/\xi_c$ for orbital pair breaking that Cooper pairs rather break at the Pauli or Clogston-Chandrasekhar limit $B_P[\text{T}] = 1.84T_c[\text{K}] \approx 2.2$ T [23,24] where the Zeeman spin splitting overcomes the superconducting condensation energy. But already preliminary experiments on $1Q1H$ [25] have shown that $B_{c2//ab}$ violates the conventional Pauli limit by an enormous factor of 10, almost as high as for the superconductors URhGe and UCoGe [26], where the ferromagnetic order hinders the Pauli pair breaking mechanism. These findings suggested the presence of an unconventional superconducting state in the $(LaSe)_{1.14}(NbSe_2)$ misfit layer, however without a proposed microscopic mechanism to explain the enormous upper critical field.

Such a strong enhancement of the in-plane critical field cannot be explained by the KLB theory taking into account spin-orbit scattering [27]. It would require unrealistically short spin-orbit scattering times given the atomic numbers $Z = 57$ and 41 for La and Nb, as compared to Ta with $Z = 73$ [20] as the spin-orbit scattering has been shown to follow the Abrikosov-Gor'kov value $\sim Z^4$ [28]. Then, different spin-orbit effects must be at play in case of $(LaSe)_{1.14}(NbSe_2)$ and $(LaSe)_{1.14}(NbSe_2)_2$. Our recent paper [4] shows that a single crystal of $1Q2H$ $(LaSe)_{1.14}(NbSe_2)_2$ is electronically equivalent to a $NbSe_2$ single atomic layer with an enormous rigid doping of 0.55–0.6 electrons per Nb atom or $\approx 6 \times 10^{14}$ cm$^{-2}$. An electronic charge transfer occurs from LaSe to $NbSe_2$ layers. Importantly, also the spin-split Fermi surfaces around $K$ and $K'$ points in the Brillouin zone of $(LaSe)_{1.14}(NbSe_2)_2$ have been observed in the quasiparticle interference (QPI) data in agreement with DFT calculations, again very similar to the case of the $NbSe_2$ monolayer. This is strongly indicating that Ising pairing is responsible for very high in-plane $B_{c2//ab}$ critical fields despite the fact that it is a fully bulk single crystal.

In the case of the $1Q1H$ $(LaSe)_{1.14}(NbSe_2)$ compound even much larger charge transfer is expected as the electronic donor LaSe is supplying not two $NbSe_2$ layers but just one. Our preliminary ARPES and STM QPI measurements [29] show a rigid band shift as compared with $1Q2H$ $(LaSe)_{1.14}(NbSe_2)_2$. Further studies are needed to determine whether the $NbSe_2$ band gets completely filled and what is the real superconducting mechanism in the system

It is worth noticing that the $NbSe_2$ surface layer of the $1Q1H$ crystal is different from the $NbSe_2$ layers located in the bulk – the surface $NbSe_2$ layers gets only half-doping compared to bulk layers which are doped from the LaSe layers laying above and underneath the $NbSe_2$ layer. Then, the electronic structure of this surface layer should be very similar to $1Q2H$. Naively, one might speculate that the superconductivity of $1Q1H$ originates from this topmost layer of $NbSe_2$ and naturally explain its low $T_c$, extreme anisotropy, 2D behavior with cusplike dependence of $B_{c2}(\theta)$ and extreme $B_{c2//ab}$. However, the specific heat measurements evidence that these effects are characteristic of the bulk of the sample. Then, in both samples the observed extreme in-plane upper critical fields are bulk effects making our $(LaSe)_{1.14}(NbSe_2)_{m=1,2}$ systems very different from the pure $NbSe_2$, where the Ising superconductivity is strong only in the case of atomic monolayer, while adding layers rapidly suppresses $B_{c2//ab}$ due to restoration of the inversion symmetry. While pristine LaSe crystals feature inversion symmetry, in $1Q1H$ and $1Q2H$ both constituents, $NbSe_2$ and LaSe, lack inversion symmetry [7,21]. If we account for the whole crystal, the total symmetry is just I1 due to incommensurability. This together with a small interlayer hopping may be responsible for Ising superconductivity in our bulk systems [30].

More recently, other mechanisms supporting large $B_{c2//ab}$ strongly exceeding $B_P$ have been discussed. In atomically thin 2D systems with inversion symmetry preserved, where the aforementioned Ising superconductivity cannot exist, a large $B_{c2//ab}$ strongly exceeding $B_P$ has been detected in $PdTe_2$ [31] and few-layer stanene [32]. A mechanism called Ising II of superconducting pairing between carriers residing in bands with different orbital indices near the Γ point is proposed, where the bands are spin-split without inversion symmetry breaking. Yet another possibility – pairing with the Rashba spin-orbit coupling – was suggested in the case of a crystalline atomic layer of In on the Si surface [33]. The particular mechanism of the extremely high in-plane critical field in $1Q1H$ remains to be explored in our future studies.

In conclusion, it has been shown that two superconducting misfit layer compounds - $(LaSe)_{1.14}(NbSe_2)$ and $(LaSe)_{1.14}(NbSe_2)_2$ with $T_c$=1.23 K and 5.7 K reveal an extremely high in-plane upper critical field reaching about 20 T and 50 T, respectively, which is 10 and 5 times more than the respective Pauli paramagnetic limit. Both compounds also show a 2D-3D crossover below $T_c$ with an upturn in $B_{c2//ab}$ temperature dependence, a very high and temperature dependent superconducting anisotropy and a cusp in the angular dependence of $B_{c2}$ for fields parallel to the layers. In $1Q2H$ $(LaSe)_{1.14}(NbSe_2)_2$ which is composed of weakly van der Waals coupled $NbSe_2$ - LaSe - $NbSe_2$ trilayers, the Ising spin-orbit coupling is most probably responsible for a very strong in-plane upper critical field. In $1Q1H$ $(LaSe)_{1.14}(NbSe_2)$, with iono-covalent bonding between LaSe and $NbSe_2$ slabs, the real mechanism behind the huge $B_{c2//ab}$ remains to be explored. Despite their possible differences a common denominator in both misfits is a strong charge transfer from LaSe to $NbSe_2$ that makes them behave as a stack of almost decoupled superconducting atomic layers with missing inversion symmetry. Further transport, STM QPI and ARPES experiments and DFT calculations are underway to elucidate superconducting mechanism in the systems.

We thank I. Mazin and M. Gmitra for valuable discussions. This work was supported by the projects APVV-20-0425, VEGA 1/0743/19, VEGA 2/0058/20, EMP-H2020 Project No. 824109, the COST action CA16218 (Nanocohybri) and VA SR ITMS2014+313011W856

---

**Supplemental Material for**
**" Extreme In-Plane Upper Critical Magnetic Fields of *H*eavily Doped**
**quasi 2D Transition Metal Dichalcogenides"**

P. Samuely, [1,2] P. Szabó,[1] J. Kačmarčík,[1] A. Meerchaut,[3] L. Cario,[3] A. G. M. Jansen[4,5], T. Cren,[6] M. Kuzmiak,[1]    O. Šofranko[1,2], T. Samuely[2]

[1]*Centre of Low Temperature Physics, Institute of Experimental Physics, Slovak Academy of Sciences, 04001 Košice, Slovakia*

[2]*Centre of Low Temperature Physics, Faculty of Science, P. J. Šafárik University, 04001 Košice, Slovakia*

[3]*Institut des Matériaux J. Rouxel, Université de Nantes and CNRS-UMR 6502, Nantes 44322, France*

[4]*Université Grenoble Alpes, CEA, Grenoble INP, IRIG, PHELIQS, F-38000 Grenoble, France*

[5]*Laboratoire National des Champs Magnétiques Intenses (LNCMI-EMFL), CNRS, UGA, F-38042 Grenoble, France*

[5]*Institut des NanoSciences de Paris, Sorbonne Université and CNRS-UMR 7588, Paris 75005, France*

**Upper critical magnetic field in (LaSe)$_{1.14}$(NbSe$_2$) and (LaSe)$_{1.14}$(NbSe$_2$)$_2$ determined at midpoint of the resistive transitions**

Figure S1 a) presents the upper critical magnetic fields of the 1*Q*1*H* (LaSe)$_{1.14}$(NbSe$_2$) misfit compound as determined at the midpoint of the resistive transition 0.5 $R_n$ of $R(B)$ from Fig. 1 c) and d) of the main text. The major features shown in Fig. 3 a) of the main text are well reproduced, here: it is an upturn of $B_{c2//ab}$, a negative derivative d$B_{c2}$/d$T$ at lower temperatures, obviously lower absolute values of $B_{c2//ab}$ pointing to about $B_{c2//ab}$ = 13 T at zero temperatures but this is still almost six times bigger than the Pauli limit $B_P$ = 2.2 T indicated by the green dashed line. Also, the temperature dependence of perpendicular field $B_{c2//c}$ of Fig. S1 a) is qualitatively similar to what is presented in Fig. 1 of the main text with a change of curvature from positive to negative one upon decrease of temperature but proportionally lower $B_{c2//c}$(0 K) = 0.35 T. The temperature dependent superconducting anisotropy $\gamma = B_{c2//ab}/B_{c2//c}$ is shown in the inset.

In the case of the 1*Q*2*H* (LaSe)$_{1.14}$(NbSe$_2$)$_2$ compound the resulting temperature dependence of $B_{c2}$ obtained from the midpoint of the resistive transitions shown in Fig. 1 e) and f) in the main text is shown in Fig. S1 b). $B_{c2//ab}$ ($T$) shows an upturn at about 4 K and $B_{c2//ab}$ reaches some 36 T at zero temperature violating $B_P$ = 10.5 T (indicated by the green dashed line) almost 3.5 times. The transversal field $B_{c2//c}$ as a function of temperature shows a positive curvature at all temperatures again in accordance with the result in Fig. 3 b) of the main text. The temperature dependence of the superconducting anisotropy $\gamma = B_{c2//ab}/B_{c2//c}$ is shown in the inset in a qualitative accord with the data in Fig. 3 of the main text.

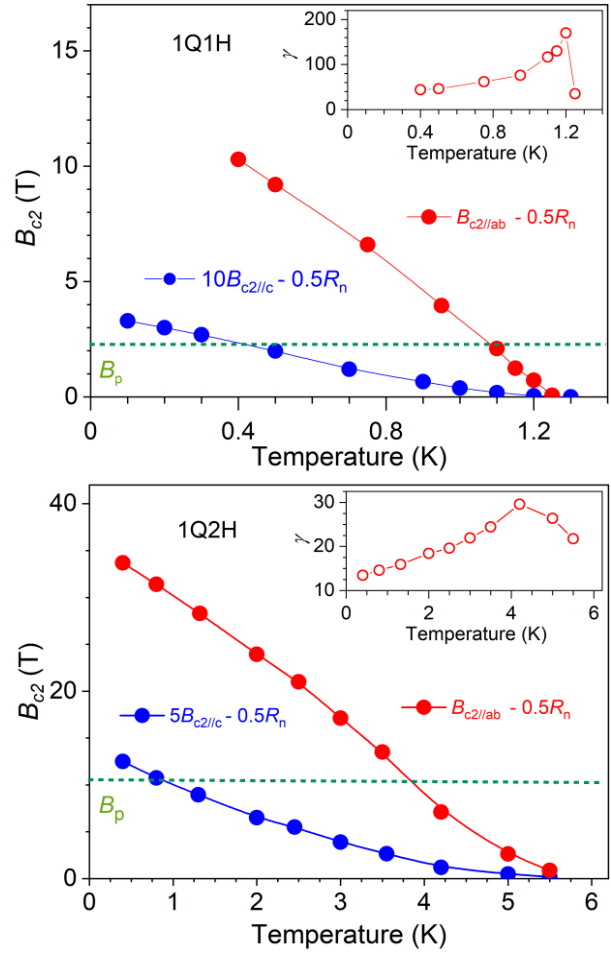

FIGURE S1. a) Upper critical magnetic fields of (LaSe)$_{1.14}$(NbSe$_2$) / 1$Q$1$H$ and b) LaSe)$_{1.14}$(NbSe$_2$)$_2$ / 1$Q$2$H$ misfit compounds determined at midpoint of the resistive transition 0.5 $R_n$. Red points – in-plane $B_{c2//ab}$, blue points - $B_{c2//c}$. Solid lines are guides for the eye. Green dashed line is the Pauli limiting field $B_P$. The temperature dependence of the superconducting anisotropy $\gamma = B_{c2//ab}/B_{c2//c}$ for both compounds are shown in the insets.

## Determination of the in-plane upper critical magnetic field $B_{c2//ab}$ from the interlayer transport measurement

Figure S2 shows the resistive transitions $R(B)$ as a function of the magnetic field oriented parallel with the ab plane of the 1$Q$1$H$ (LaSe)$_{1.14}$(NbSe$_2$) misfit compound at fixed temperatures. While in the part a) the measurements with the current and voltage electrodes put on the same top surface *ab* plane reproduce the intralayer resistance $R_{ab}$ what is already presented in Fig. 1 d) of the main text, in Fig. S2 b) we display interlayer measurements of $R_c$ taken at configuration where one voltage and one current contact are put on the top *ab*-surface plane and the other pair of contact electrodes are on the bottom *ab*-surface plane of the sample. At all indicated temperatures upon increased magnetic field *B* the zero-resistant state is followed by the onset of finite resistance increasing to a peak, then decreasing to an almost constant background with a slightly positive magnetoresistance in the normal state. All the data have been normalized to the value at 20 T. At decreasing temperatures the peak is broadened, its amplitude suppressed and position shifted to higher fields. The peak position is always found close to the field where the intralayer resistance $R_{ab}$ drops to zero. Figure S2 c) displays the interlayer magnetotransport data recalculated in the conductance, $\sigma_c = 1/R_c$. One can see that before reaching the normal state a linear dependence of the applied field is present, which is stressed by the dashed lines. The intersections of the dashed lines with the normal state background have been taken for determination of the upper limit of the in-plane upper critical magnetic field $B_{c2//ab}$. This is presented in the Fig. 3 a) of the main text. In our previous work [9] we have found that the highly anisotropic $(LaSe)_{1.14}(NbSe_2)$ misfit layer compound behaves as a stack of intrinsic Josephson junctions. Peak in the interlayer resistance $R_c$ preceding a full transition to the superconducting state, there measured at fields parallel to the *c* axis of the crystal, was interpreted as due to the interplay between the quasiparticle and Josephson tunneling across the atomic layers creating intrinsic Josephson junctions. At small fields first, Josephson coupling between superconducting planes is suppressed, finite resistance/conductance appears with a maximum/minimum. At increasing fields the number of the vortices is increased proportionally to the field strength, cores of the vortices are normal areas with quasiparticles allowing for quasiparticle tunneling between decoupled yet superconducting layers causing a decrease of the resistance and increase of the conductance until the full normal state is achieved.
.

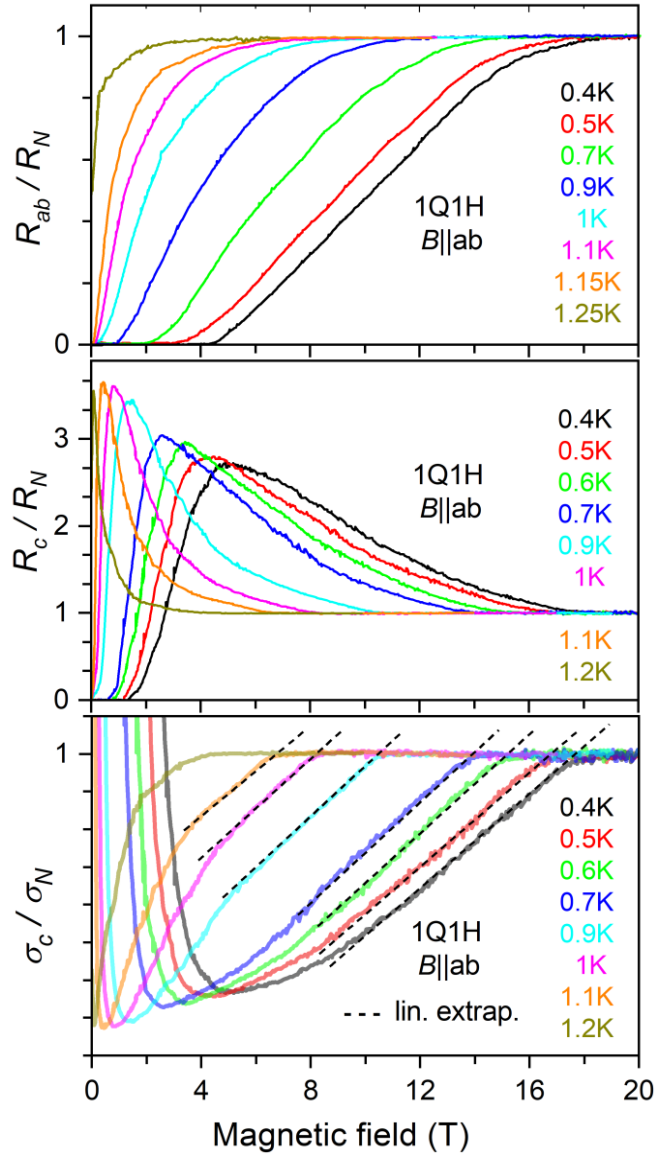

FIGURE S2. a) In-plane resistive transitions $R_{ab}$ as a function of magnetic field oriented parallel to the *ab* plane of the $(LaSe)_{1.14}(NbSe_2)$ / $1Q1H$ misfit compound at fixed temperatures (same as in Fig. 1 d) of the main text). b) Interlayer measurements of $R_c$ taken at configuration where one voltage and one current contact are put on the top *ab*-surface plane and the other pair of contact electrodes are on the bottom *ab*-surface plane of the sample. c) displays the interlayer magnetotransport data recalculated in the conductance, $\sigma_c = 1/R_c$. Dashed line is linear extrapolation. All resistances are normalized to the high field value.

**Angular dependence of $B_{c2}$**

Figure S3 displays the experimental data of the resistance of the $1Q2H$ $(LaSe)_{1.14}(NbSe_2)_2$ misfit compound as a function of applied magnetic field oriented at different angles $\theta$ with respect to the *ab* plane of the crystal with $\theta = 0$ for magnetic field parallel to the *ab* plane. The data has been used to construct the angular dependence of the upper critical magnetic field $B_{c2}$ presented in Fig. 1 h) of the main text.

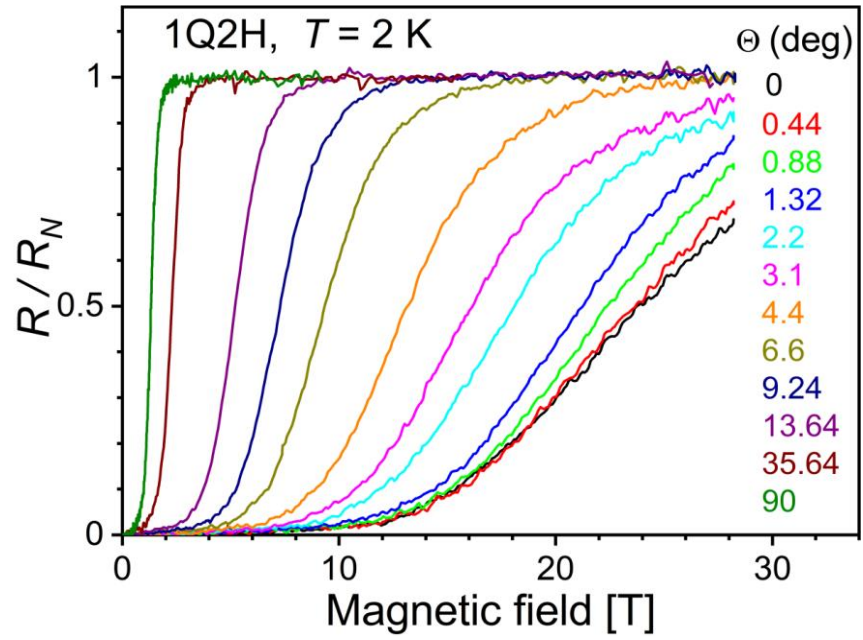

FIGURE S3. Resistive transitions $R$ as a function of magnetic field oriented at different angle $\theta$ with respect to the *ab* plane of the sample. The data is taken to determine the angular dependence of $B_{c2}(\theta)$ in Fig. 4 b) of the main text.